\documentstyle [preprint,prl,aps,epsf]{revtex}
\begin {document}

\title {Comment on ``Quantum Confinement and Optical Gaps 
in Si Nanocrystals''}
\author {Alberto Franceschetti, Lin-Wang Wang, and Alex Zunger}
\address {National Renewable Energy Laboratory, Golden, CO 80401.}
\maketitle

\vskip 1.2cm

In a recent Letter \cite {OCL} \"O\u g\"ut, Chelikowsky,
and Louie (OCL) calculated the optical gap
of Si nanocrystals as
$\varepsilon_{g,OCL}^{opt} = \varepsilon_{g,LDA}^{qp} - E_{Coul}^{e-h}$,
where $\varepsilon_{g,LDA}^{qp}$ is the quasi-particle 
gap in the local-density approximation (LDA)
and $E_{Coul}^{e-h}$ is the electron-hole Coulomb energy.
They argued that their method produces different results
from conventional approaches (e.g. pseudopotential \cite {EPM}).
We show in this Comment that the expression of  
$\varepsilon_{g,OCL}^{opt}$ 
omits an electron-hole polarization energy
$E_{pol}^{e-h}$. When this contribution is taken into account
(together with the LDA correction $\Delta_{LDA}$ 
pointed out in Ref. \cite {Godby})
the optical gap $\varepsilon_g^{opt} = 
\varepsilon_{g,LDA}^{qp} + \Delta_{LDA} - 
(E_{Coul}^{e-h} + E_{pol}^{e-h})$
is in excellent agreement with the pseudopotential approach \cite {EPM}.

The quasi-particle gap of a neutral, $n$-electron nanocrystal
can be obtained as 
$\varepsilon_{g}^{qp} = [E(n+1) - E(n)] + [E(n-1) - E(n)]$,
where $E(m)$ is the total energy of the $m$-electron system \cite {OCL}.
Following Ref. \cite {Brus},
the dependence of $E(n+1) - E(n)$ and $E(n-1) - E(n)$ on the
nanocrystal size can be derived from classical 
electrostatic considerations,
and the quasi-particle gap can be written as
$\varepsilon_g^{qp} = \varepsilon_g + \Sigma_{pol}$,
where $\varepsilon_g$ is the single-particle gap
and $\Sigma_{pol}$ is the polarization energy
originating from the interaction with the surface 
polarization charge \cite {Brus}.
$\Sigma_{pol}$ can be evaluated using 
effective-mass envelope functions:

$$\Sigma_{pol} \simeq
{e^2 \over R} \left [
{1 \over \epsilon_{out}} - {1 \over \epsilon_{in}}
+ {0.94 \over \epsilon_{in}} 
\left ( {\epsilon_{in}  - \epsilon_{out} \over
\epsilon_{in} + \epsilon_{out}} \right ) \right ] \; , \eqno (1)$$

\noindent where $\epsilon_{in}$ is the {\it size-dependent}
dielectric constant of the nanocrystal, $\epsilon_{out}$
is the dielectric constant of the barrier (i.e. vacuum),
and $R$ is the nanocrystal radius.
We have calculated $\Sigma_{pol}$ from Eq. (1) using
the dielectric constant $\epsilon_{in}$ of OCL.
As shown in Fig. 1(a), the self-energy correction
$\Sigma_{OCL} = \varepsilon_{g}^{qp} - \varepsilon_{g}$ 
of Ref. \cite {OCL} is almost entirely due to the 
classical polarization energy $\Sigma_{pol}$.

Having established that $\varepsilon_{g}^{qp}$
includes the polarization effect,
we discuss now how to derive the optical gap
from the quasi-particle gap $\varepsilon_{g}^{qp}$.
Following Ref. \cite {Brus},
in addition to the {\it direct} electron-hole
Coulomb energy $E_{Coul}^{e-h}$
one should include (in analogy with $\Sigma_{pol}$ above)
the interaction between the electron and the surface polarization
charge produced by the hole, 
and the interaction between the hole and the surface 
polarization charge produced by the electron.
We have calculated this polarization energy with the same
approximations used in the calculation of $\Sigma_{pol}$. We obtain:

$$E_{pol}^{e-h} \simeq
{e^2 \over R} \left (
{1 \over \epsilon_{out}} - {1 \over \epsilon_{in}} \right )
\; . \eqno (2)$$

\noindent We see from Fig. 1(a) (dashed line) that $E_{pol}^{e-h}$
is comparable in magnitude with $\Sigma_{pol}$.
This term was incorrectly neglected in Ref. \onlinecite {OCL}.
Also, since the LDA quasi-particle gap $\varepsilon_{g,LDA}^{qp} (R)$ 
does not have the correct $R \rightarrow \infty$ limit \cite {Godby},
an LDA gap correction $\Delta_{LDA}$ 
needs to be added \cite {Godby,Delley}. 
Thus, the correct optical gap is
$\varepsilon_g^{opt} = \varepsilon_{g,OCL}^{opt} - 
E_{pol}^{e-h} + \Delta_{LDA}$.

If we define 
${\bar \varepsilon}_g \equiv \varepsilon_g^{opt} + E_{Coul}^{e-h}$,
we obtain from the above equations
${\bar \varepsilon}_g = \varepsilon_g + (\Sigma_{pol} - E_{pol}^{e-h})$,
where
$\varepsilon_g = \varepsilon_{g,LDA} + \Delta_{LDA}$
is the (LDA-corrected) single-particle gap and 
$\Sigma_{pol} - E_{pol}^{e-h}$ is the net polarization energy
which is small [Fig. 1(a)].
In standard models of excitons in nanostructures it is assumed
that ${\bar \varepsilon}_g \simeq \varepsilon_{g}$,
and the single-particle gap $\varepsilon_g$ is 
obtained e.g. with $\bf k \cdot p$ or pseudopotentials (ps).
Fig. 1(b) compares ${\bar \varepsilon}_g$ 
with $\varepsilon_{g,ps}$ \cite {EPM}.
The excellent agreement suggests that
the optical gap of Ref. \onlinecite {OCL},
when corrected for both the polarization error ($E_{pol}^{e-h}$)
and the LDA error ($\Delta_{LDA}$),
is consistent with the conventional optical gap.

\begin {figure}
\vskip 1.0cm
\centerline {\epsffile {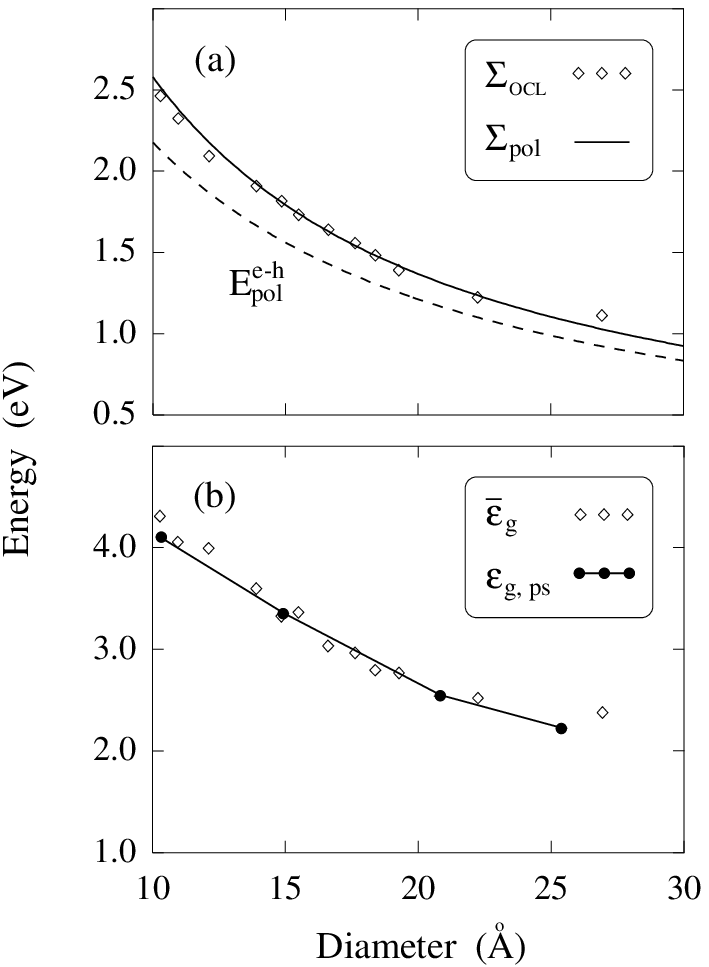}}
\vskip 1.0cm
\caption {The polarization energy $\Sigma_{pol}$
is compared in part (a) with the self-energy $\Sigma_{OCL}$ [1].
Part (b) compares $\varepsilon_{g,ps}$
with ${\bar \varepsilon}_g$ (see text). The LDA correction
$\Delta_{LDA}$ is assumed to be constant (0.68 eV) for 
$D > 10 \, {\rm \AA}$.}
\end {figure}

\end {document}